# A Cloud Security Framework Based on Trust Model and Mobile Agent

Saddek Benabied
LIRE Labs
University of Constantine 2
Constantine, Algeria
benabied_s2005@yahoo.fr

Abdelhafid Zitouni
LIRE Labs
University of Constantine 2
Constantine, Algeria
ah_zitouni@yahoo.fr

Mahieddine Djoudi
TechNE & XLIM-SIC Labs
University of Poitiers
Poitiers, France
mdjoudi@univ-poitiers.fr

*Abstract*— Cloud computing as a potential paradigm offers tremendous advantages to enterprises. With the cloud computing, the market's entrance time is reduced, computing capabilities is augmented and computing power is really limitless. Usually, to use the full power of cloud computing, cloud users has to rely on external cloud service provider for managing their data. Nevertheless, the management of data and services are probably not fully trustworthy. Hence, data owners are uncomfortable to place their sensitive data outside their own system .i.e., in the cloud.

Bringing transparency, trustworthiness and security in the cloud model, in order to fulfill client's requirements are still ongoing. To achieve this goal, our paper introduces two levels security framework: Cloud Service Provider (CSP) and Cloud Service User (CSU). Each level is responsible for a particular task of the security. The CSU level includes a proxy agent and a trust agent, dealing with the first verification. Then a second verification is performed at the CSP level. The framework incorporates a trust model to monitor users' behaviors. The use of mobile agents will exploit their intrinsic features such as mobility, deliberate localization and secure communication channel provision. This model aims to protect user's sensitive information from other internal or external users and hackers. Moreover, it can detect policy breaches, where the users are notified in order to take necessary actions when malicious access or malicious activity would occur.

## I. INTRODUCTION

During the last years, processing and storage technologies procure rapid development, decreasing costs, and increasing power of the computer resources beside to the success of the Internet, led to a situation where a gigantic quantity of data can be collected, stored and reached.

In that context, cloud computing is rapidly appeared as IT (information technology) solution of choice for many organizations and individuals. Because cloud architecture relies on distant rather than local servers, it provides dynamically scalable infrastructure or virtualized resources, platform and software in the form of services over the Internet. It is enabling the realization of a new computing model, it proposes a transition from the traditional economic model in which the user is the owner of the software and the hardware, towards a model in which the user is becoming a simple tenant of services, which they are software or even hardware, and they are provided as general utilities that can be leased and released by users through the Internet in an on-demand fashion and pay as you go model. In a cloud computing environment, the traditional role of service provider is divided into two: the *infrastructure providers* who manage cloud platforms and lease resources according to a usage-based pricing model, and *service providers*, who rent resources from one or many infrastructure providers to serve the end users [4].

Due to cloud's computing deployment characteristics, as multi-tenancy nature; the outsourcing of sensitive data; the critical applications and infrastructure in cloud is troublesome. The organizations and individuals are very anxious about how security can be guaranteed in the new cloud environment. Moreover, enterprises and individuals have strong constraints on hosting their sensitive data and critical applications on clouds.

So, the biggest challenge in cloud computing is to successfully address the security issues associated with their deployment [13, 15], and to provide evidence to their customers that their data are safe.

In order to alleviate these security fears, and to add a brick to cloud computing security research, we propose in this paper a new architecture to implement security. It is based on continuous examination of trust levels. It is a two tier framework based on trust model and mobile agents. 1) Trust model calculates users' trust with respect to their domain trust degree. 2) Monitor users' behaviors. The use of mobile agents will exploit their intrinsic features such as mobility, deliberate localization and secure communication channel provision. Thus, replacing client-server model usually employed to ensure that customers will experience the same security and privacy will reinforce the trust between cloud service providers and user.

The rest of this paper is organized as follows: In section 2, we provide a review for the cloud computing domain. Section 3 provides a survey of cloud security, threats and challenges. Section 4 illustrates some related works. Section 5 proposes our framework and model for cloud security. Finally, section 6 gives our conclusion and future work.

## II. CLOUD COMPUTING OVERVIEW

This section presents a general overview of cloud computing, including its definition, deployment models, services, its characteristics and a comparison with related concepts.

### A. Definitions

For the companies and end-users (from the multinational large account at local Startups), Cloud Computing can be defined like an approach allowing them to have applications, computing capacity, storage tools, etc. as much of services . Those will be shared, dematerialized, centralized, evolutionary (in volume, function, characteristic.) and in self-service. [01] [02]

In this paper, we adopt the definition of cloud computing provided by The National Institute of Standards and Technology (NIST) ,as is being based on the combination of various criteria, and it covers all the essential aspects of cloud computing. The NIST working definition summarizes cloud computing as:

Cloud computing is a model for enabling convenient, on-demand network access to a shared pool of configurable computing resources (e.g., networks, servers, storage, applications, and services) that can be rapidly provisioned and released with minimal management effort or service provider interaction [03].

The essential characteristics of cloud computing are:

- **On-demand self-service:** computing resources can be acquired and used at anytime without the need for human interaction.

- **Broad network access:** the computing resources can be accessed over a network using heterogeneous devices.

- **Multi-tenancy:** the different resources can be shared by multiple users.

- **Rapid elasticity:** a user can quickly acquire more resources from the cloud by scaling out. They can scale back in by releasing those resources once they are no longer required.

- **Measured service:** resource usage is metered using appropriate metrics such monitoring storage usage, CPU hours, bandwidth usage etc.

Cloud computing provides in general three types of service (Figure 1).

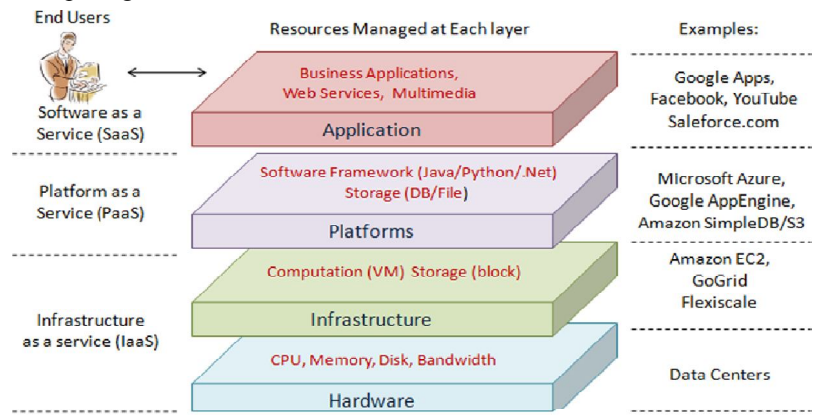

Figure 1. Cloud computing architecture [4].

## III. CLOUD COMPUTING AND SECURITY CHALLENGES

In the recent year, the cloud computing is emerged and has been adopted by many IT organizations. This paradigm has changed the overall architecture and system requirements, compared to traditional server-based systems. Cloud-based system architecture provides Internet-based services, computing and storage in all fields including health care, finance, government etc with the reduced price [10]. Despite the benefits offered by cloud computing, many companies still have apprehensions about them. This is most likely due to the various security issues that have yet to be dealt with. There is a serious concern for an organization to move towards the cloud, where access is ubiquitous, information exchange is abundant and data location is often unknown [6]. The majority of the companies mention the data protection as their major security issue.

The companies must have the certainty that, on the one hand, the authorized users within the company have access to the necessary information and tools which they need, when they need them, and on the other hand, the not - authorized accesses are blocked. These controls are more essential when the Cloud environments which generally used by a wide and diversified community of users. Moreover, cloud computing introduces a new category of privileged users, the administrators who work for the Cloud service provider. Monitoring of these particular users, and specially the journalizing of their activities, has a great importance. Although cloud computing emerges from existing technologies, its computing (delivery and deployment) models and characteristics raise new security challenges due to some incompatibility issues with existing security solutions [11].

In order to have a secured Cloud computing, many areas must be considered as: the cloud computing architecture, Governance, portability and interoperability, traditional security, business continuity and disaster recovery, data center operations, incident response, notification and remediation, Application Security, Securing Data Storage, Encryption and Key management, identity and access management [05].

*A. Information Security*

The term information security means protecting information and information systems against unauthorized access, use, disclosure, disruption, modification, or destruction, to provide confidence, integrity and availability for the users. The basic security issues are: authentication, authorization, auditing, confidentiality, integrity, availability and non-repudiation [11], [06]:

*Authentication:*

Authentication is the process by which a user or other entity's identity is verified.

*Authorization:*

Authorization is the process that ensures that an entity has the right credentials to access certain resources or to perform a requested task. This decision is made after authenticating the identity in question. And each entity has a specific level of authorization.

*Confidentiality:*

Confidentiality is to protect personal privacy and proprietary information, which should not be read by others as unauthorized individuals, entities or processes.

*Integrity:*

Integrity is ensuring the data protection against improper modification or destruction by unauthorized persons or processes during the course of transmission.

*Non-Repudiation:*

Non-repudiation is the ability to limit entities from refuting that a legitimate transaction took place, and to ensure that it has not been changed by a malicious entity.

*Availability*:

Is the ability to ensure that data or resources are available at any time in continuous manner when and where are needed.

*Auditing:*

Is a process of recording of the performed tasks and collecting information about user attempting access to a particular resource?

*B. Threats of cloud computing*

As more companies move to cloud computing, due to its characteristics, hackers also follow this way.

*Threats to cloud computing discovered by "Cloud Security Alliance" (CSA):* Cloud Security Alliance has proposed the biggest security threats of cloud systems as follow [12]:

*a) Abuse and Nefarious Use of Cloud Computing:* hackers use the power of the cloud infrastructure to attack other machines. It was ranked as the top threat identified by the CSA [12].

*b) Insecure Application Programming Interfaces:* The cloud's costumers interact with cloud services via APIs; those must be extremely secured.

*c) Malicious Insiders:* It is the threat when an authorized cloud user gains from his authority and access to another user account or resources.

*d) Shared Technology Vulnerabilities:* Infrastructure providers rely on sharing infrastructure technology for offering their services. Unfortunately, the components on which this infrastructure is based were not designed for that.

*e) Data Loss/Leakage:* It can be done by deletion without backup, losing the encoding key for encrypted data or by unauthorized access.

*f) Account, Service & Traffic Hijacking:* These threats range from man-in-the-middle attacks, to phishing and spam campaigns, to denial-of service attacks.

1) *Security problems concerning Location, Network Concept and virtualization:* they are inherited from cloud characteristics [6], [13]:

*a) Multi-location of data:* In cloud computing data is stored anywhere around the world and users can't know anything about it location.

*b) Denial of Service (DoS) attacks:* Some security professionals have argued that the cloud is more vulnerable to DoS attacks, because it is shared by many users, which makes DoS attacks much more damaging.

*c) Side channel attacks:* An attacker could attempt to compromise the cloud by placing a malicious virtual machine in close proximity to a target cloud server and then launching a side channel attack.

*d) Authentication attacks:* Authentication is a weak point in hosted and virtual services and is frequently targeted.

*e) Man in the middle attacks:* an attacker intrudes himself in an ongoing communication between two users, and trying to inject false information or have knowledge of data transferred between them.

*f) Inside job:* This kind of attack is when the person, employee or a staff who is knowledgeable of how the system runs, from client to server then he can implant malicious codes to destroy everything in the cloud system.

## IV. RELATED WORKS

Over the last decade a number of researchers used agents in their works, that in theory could provide some basic levels of computer and network defense that could be applied to Cloud computing security.

In [8] Shantanu & *al*, proposed a cloud security two-tier framework based on a collaborative agent and a trust model to protect cloud resources by monitoring the unauthorized access. They proposed a novel trust model assures the trustworthiness of the system by calculating current or updated trust degree for each user service request and the domain from where the request is coming. In their model a proxy server is used as a communication channel between two domains and to authenticate each service request and also to deliver the response to the service user. In the two levels cloud users and cloud provider, two agents are used to calculate and update the user and domain trust degree using the trust function [7], also they maintain their databases and user activities. So, only the users who have a trust degree greater than a threshold trust can access and get the information from cloud service provider. But malicious user can't access and they will be removed from its domain.

This model has a major advantage that the domain remains unaffected by non trusted users (with only decreased amount of trust degree). But this model suffers from some weakness, because it increases some workload of domains and it fails to prevent malicious activity without cloud service provider information about user activities, and the proxy server presents also a weak point, if it crashes, the users can't communicate with the cloud service provider.

In [09] Priyank & *al*, proposed a trust model agent based technique for cloud architecture. Mobile agents are used as security agents to collect information from the virtual machines in order to help the user and service provider for keeping track of privacy of their data and virtual machines. These agents monitor virtual machines integrity and authenticity. Security agents can circulate and migrate in the network, replicate and perform the assigned tasks for monitoring of virtual machines. Mobile agents are introduced at multiple levels in the cloud infrastructure in order to monitor the resources utilization and minimize security threats; also they used security agents to build a strong trust between cloud users and cloud providers. But they ignored the user identification, the identity management, and the security of the agents themselves.

In [10] Amir et al, investigated the problem of data security in cloud data storage, and in order to verify the security and availability of users' data in the cloud, they proposed a multi agent system framework. The proposed framework has been built by using two layers: agent layer and cloud data storage layer. The SMA built by six agents, each one has a specific task and they communicate to achieve the global goal, the tasks are: User Interface assistance of user in operating an interactive interface; tolerate multiple failures in distributed storage systems, enable the user to reconstruct the original data by downloading the data vectors from the servers, storing associated information and storing system information, as well as recording the messages and data shared among agents.

In [16] authors proposed a multi agent system framework, it presented as a two layer framework: a virtual server composed by a set of agents to intercept the users request, and the cloud layer also contain a set of agents to perform the requested service in the cloud, each agent has a specific task to do. They used a mobile agent as communication entity between the two layers, which is benefic for reducing the traffic on the network and the amount of the information exchanged. They presented their methods as a multi agent system cooperating to achieve a global goal which is satisfying user's service request, but they don't clarified how the security of the system is maintained.

V. THE PROPOSED FRAMEWORK

*A. The Proposed Framework Architecture*

The system architecture is presented in Figure 6. This is composed of two tier; cloud service provider and cloud service user. The framework is relying on a proposed novel monitoring techniques assures the trustworthiness of the system by calculating current or updated trust degree for each CSU, who request a service from CSP and the domain from where the request is coming, and a set of agents ensuring a secure data communication between different stakeholders, that is done by agent's mobility. A set of agents is used in the proposed architecture, and they are described below with their functions:

1) *Interface Agent:* This agent gives a graphical interface and helps the users to interact with the system; it's a static agent that runs on user's device. Its functions are: it is responsible for acquiring user's requests, sending them to appropriate agent and showing results to users. Its architecture contains different modules as shown in (Figure 2):

- Interface module: it allows the user to interact, to request and to visualize the results;

- Treatment module: to analyze the user's data and collecting information from the graphical interface and sends the request to the proxy agent,

- Security communication module: for securing the communication between this agent and proxy agent;

- Knowledge base: it stores the agent's knowledge about the information transmitted by the interface agent and the profile of the current user.

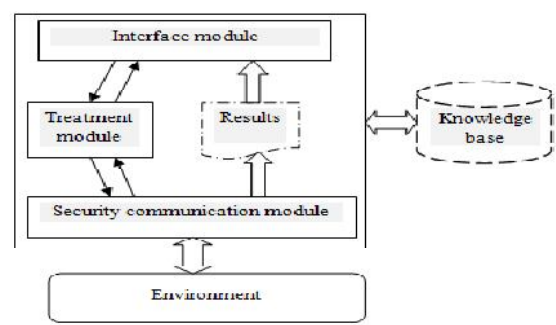

Figure 2. Interface agent.

2) *Proxy Agent:* The interface agent communicates to the proxy agent the authentication information, the proxy agent analyzes this information to authenticate users, and it communicates with trust user agent to verify the user trust degree. Also it is responsible for creating the communication channel between the two tiers by activating a mobile agent and sends it to the cloud service provider site. It conveys response to the appropriate interface agent (Figure 3):

- Access management module: this module is responsible for checking the authentication of each user and their credentials;

- Data base: in this base is stored all the knowledge of the proxy agent about: users and their credentials, names and address of the deferent agents communicating with it;

- Security communication module: is responsible for securing the communication between this agent and the other agents;

- Treatment module: it is responsible to manage the user's data, collecting information from access authorization and keep user traces.

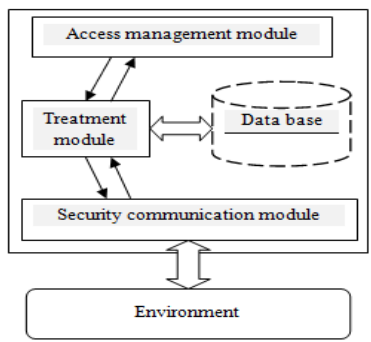

Figure 3.   Proxy agent.

*2) Mobile Agent:* The mobile agent is launched by the proxy agent and it migrates to a cloud service provider site as missionary. Its functions are: carrying the user request and their results, to ensure a secure communication channel between the two stakeholders, the interaction with trust agent, and the interaction with the cloud service provider. Finally, he destroys himself:

- Treatment module: it is responsible to carry the user's request and result to or from cloud service provider, and it communicates with trust agent to check user domain trust;

- Secure communication module: for security interaction between this agent and the other agents.

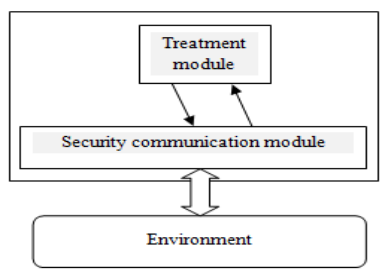

Figure 4.   Mobile agent architecture

*3) Trust Agent:* for checking and updating the user trust degree according to the trust algorithm and user activities (Figure 5):

- Secure communication module: is responsible for security interaction between this agent and the other agents;

- Treatment module: for evaluating user trust and his behavior and updating it.

- Knowledge base: for storing the trust agent knowledge about user with his trust and its evolution.

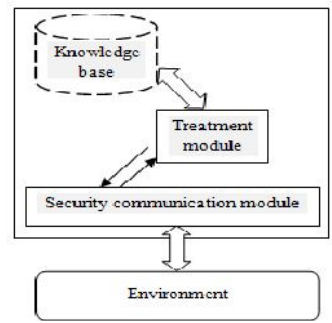

Figure 5.   Trust agent architecture.

*B. Framework description*

In a heterogeneous cloud computing environment it is very important to ensure the two cloud issues: firstly, the security of data communication, because data and software are stored, accessed and run on machines that are not owned or directly managed by owners of data and software, secondly, the trust between Cloud Service Provider (CSP) and Cloud Service User (CSU). Agent-based models and algorithms for trust and security in Cloud infrastructures could be very useful to determine and satisfy the basic trust requirements of the service requesting CSUs.

The agent's technology offers various solutions using agents to cloud vulnerabilities as: control access and authentication, distributed trust management, audit and intrusion detection, attack vector pursuit, and diagnostic and system restoration. Autonomous agents can make Clouds smarter in the interaction with users and more efficient in allocating processing, resources and storage to applications and users.

In Clouds, there is a veritable need to design and implement techniques that can monitor the dynamic behaviors, changing configurations and heterogeneity of Cloud computing environments. Autonomic techniques as multi agent systems appear suitable to provide a promising approach for addressing this requirement.

For these reasons and for others that we have discussed above, two-tier architecture based on mobile agents has been proposed in this paper. These two level communications authenticates the CSUs and monitors their trust and behaviors for accessing and handling private information from cloud data storage through a set of agents.

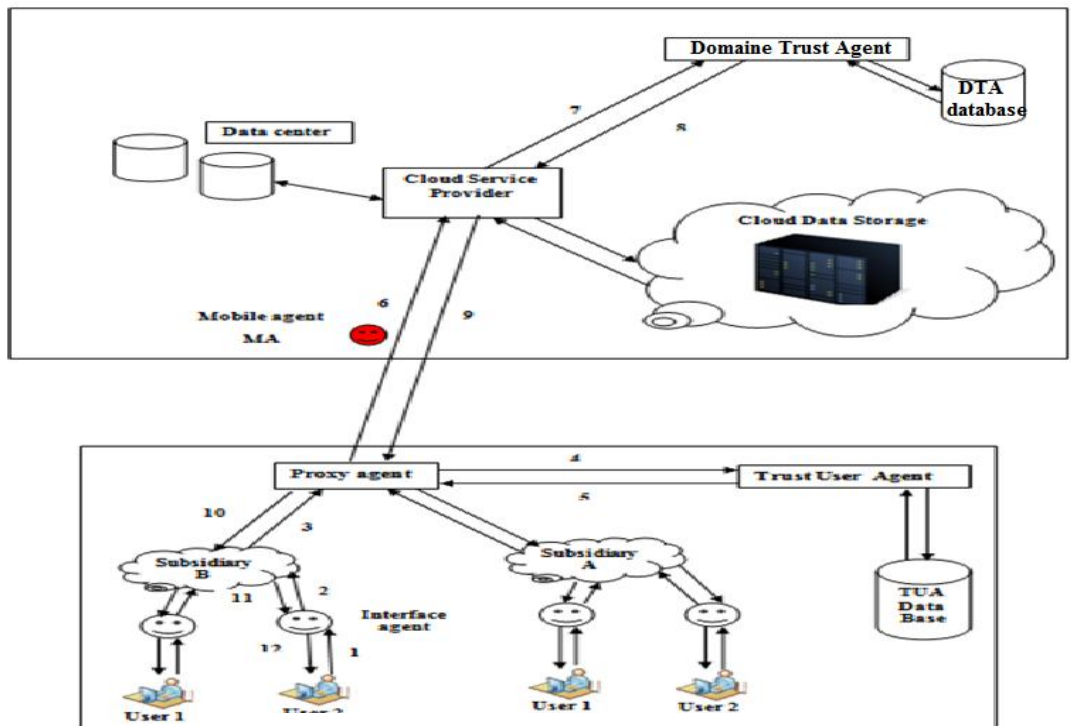

Figure 6. Proposed Model

In this framework the CSU (user1, user2… as depicted in Figure 6) requests a service from the Cloud Service Provider (CSP). As the first step: authentication, when any user who wants to use any service of the CSP, he has to pass the correct authentication information (such as id and password) via the interface agent to the proxy agent situated in his domain (for example his subsidiary). In this approach the proxy agent is used as a communication channel between domains and a manager for sending and receiving user's request to and from CSP. As an example, subsidiary A and subsidiary B is denoted for a specific group of users. When the request arrives to the proxy agent, the proxy agent asks the trusted-agent (Trusted User Agent TUA in Figure 6) about the trustworthiness of the user who sends this request. Then the TUA regarding its database checks the trust level of the CSU and replies to the proxy agent request. Then only if the trust level of this CSU is greater than a predefined threshold degree, the proxy agent launches a mobile agent (denoted as MA) who migrates through internet to the Cloud Service Provider host. When the MA arrives to the CSP host, it sends immediately a request to the Domain Trust Agent (denoted as DTA) to check the trustworthiness of the CSU and the domain from where this request has came. The Trust Agent DTA then checks the current or updated (this updating is done regarding the previous activities) trust level for this particular domain and then sends this result back to the MA, the MA in his turn communicates all of these information (user request, user trust degree, domain trust degree) to the CSP. Only if the trust level is greater than the predefined threshold degree set for this domain, then the CSP will allow MA to get the requested information and to transmit it back to the particular CSU through the proxy agent.

The DTA is responsible for monitoring and updating the trust level after each performed task within the system. If the domain trust level is less than the threshold level, or the CSU does any malicious activities, the DTA immediately informs the TUA at the respective domain to take the necessary actions. After receiving the information from DTA the TUA will proceed for decreasing the trust level for this particular user. In the case of a repetitive non-trusted activities or reports of a malicious or non trusted user, the TUA will remove this particular user from his domain. The whole procedure is detailed according to the sequence diagram in following steps (Figure 7):

- STEP1: the user sends his identification information to the interface agent.
- STEP 2: the interface agent sends the identification information to the proxy agent.
- STEP 3: the proxy agent checks the credentials and if the access information is correct, it establishes a secure SSL connection with interface agent, and asks it to load the interface, and to transmit the user request. Otherwise the access request is rejected.
- STEP 4: the proxy agent sends the user identifiers' to the trust user agent (TUA) for trust level checking.
- STEP 5: trust user agent calculates the trust level of the user, and compares it with a specific threshold, if it is greater than this threshold TUA returns response(trusted user) to proxy agent, otherwise returns(not trusted user).

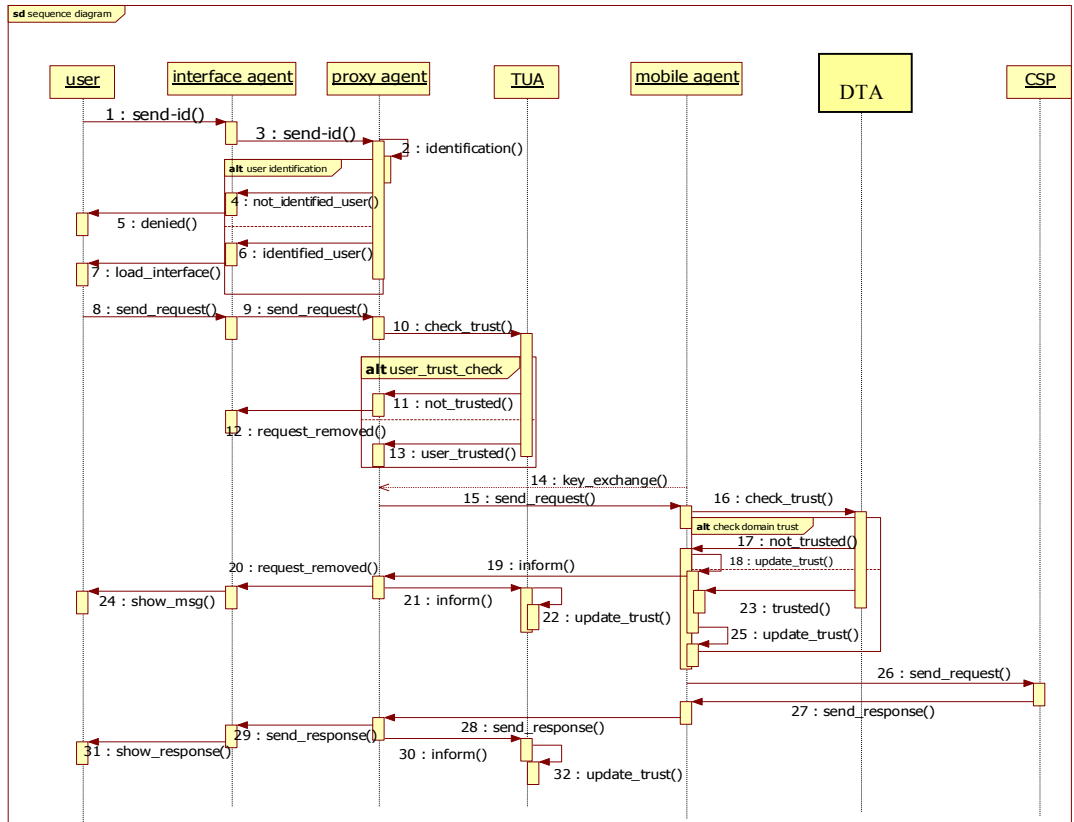

Figure 7. Sequence diagram

- STEP 6: the proxy agent receives the TUA response, if the response is non trusted user, the request will be removed, else, the proxy agent creates a mobile agent (MA), and sends it to the cloud service provider site, carrying with it user's, domain's identification and the encrypted request.
- STEP 7: the MA when arrived at CSP's site, it establishes a secure SSL connection with Proxy Agent and sends a request with user's and his domain identifications to the domain trust agent.
- STEP 8: the DTA evaluates the trust level of user domain, and compares it with a specific threshold set for this domain, if it is greater than this threshold, the DTA replies with (trusted domain) to the mobile agent, otherwise returns(not trusted domain).
- STEP 9:

a) According to the response of DTA: if non trusted domain, the MA deletes the request and informs the proxy agent, which also informs the TUA for further actions. Otherwise, if trusted domain, the MA communicates the request with CSP.

b) The CSP provides service to the MA .

c) The MA communicates the information with the proxy agent.

- STEP 11: the proxy agent receives the information and transmits it to the right interface agent and informs TUA for updating user trust level.
- STEP 12: interface agent displays the information to the user.

### C. algorithm for calculating Trust level:

The proposed security model is based on continuous examination of trust levels, a trust algorithm and a set of agents are used to determine and monitor the users trust level. The trust level of each CSU can increases or decreases depending on his behavior, i.e., when performing any trusted or no trusted task. The trust agents (TUA and DTA) calculate the users trust degree and update their own databases using the trust algorithm [7] explained below.

Each trust agent uses a trust level function to calculate the trust level of each CSU requesting a service from the CSP, and after calculating of the trust level, the users will be classified as trusted, innocent or non-trusted user. With the use of that function the probability to have a good trust value for a trusted user performing a trusted task, is always higher than a non-trusted or an innocent user trust level [7].

We can classify the Actions for any task into two sets positive or negative actions. The Positive actions are the right actions performed by the trusted user; the negative actions are the bad actions done by any user. However, it is clear that all negative actions are not the same. So, the negative actions are divided on wrong actions and malicious actions [7]: Wrong, i.e. negative actions that do not cause any damage or may cause damages done by the innocent user; and Malicious, i.e. negative actions that cause damages such as attacks done by the non-trusted user.

To calculate the probability to do a positive action $P_a$, we take into account the user behavior and the performed action weight. This function increases or decreases according respectively to the performed positive and negative actions. So, the function is denoted as [7]:

$$Pa = \left(1 - \frac{N_a}{total_a}\right) \cdot W_a^{(l)} \qquad (1)$$

Where $0 \leq Pa \leq 1$

$N_a$ : is the number of negative actions;

$total_a$: is the total number of performed actions;

*Wa:* is the action's weight according to its nature (positive, wrong, and malicious) depending upon the requesting cloud service user and performance of task ($0 \leq wa \leq 1$). Parameter *l* is the security level, $l >= 1$.

In (1), $1 - \frac{N_a}{total_a}$ : represents the past behavior of any CSU. This value tends to 0 when the behavior is malicious, and it tends to 1 when the behavior is trusty.

The objective is to minimize the probability of malicious users to gain access to the service, for this reason we raise the action weight to the power of (*l*). This security level affects the action weight. The exponential really influences when the actions are wrong. When the user performs a new action, *Pa* is recalculated, reflecting the present behavior of the CSU. The trust agent will take it into account and modify the current trust value of the service requesting CSU.

The major advantages of this framework are: it reduces the traffic on the network and decreases the quantity of information exchanged between the two levels users and the cloud provider. Also, the trust level of the user will decrease or increase accordingly to his activities (malicious or not), so, that can help for detecting policy breaches and updating policies. The two trust agents: DTA and TUA maintain their own trust databases especially: user activities information and old trust levels for calculating updated trust level.

## VI. CONCLUSION AND FUTURE WORKS

In this paper, we have proposed a two levels security framework based on continuous examination of trust degree, a trust model is used to calculate the trust degree of each user, and to monitor his behavior and activities. It is beneficial for updating security policies, to prevent unauthorized accesses to cloud data and for protecting information against malicious users and misuse. It is based also, on mobile agent's technology and their features for cloud computing security. The advantage of this architecture is that it uses mobile agents as a communication entity. This is to reduce traffic on the network, to reduce work load at cloud service provider, to reduce the amount of information exchanged between user and CSP, so, it minimizes the chance to intercept the exchanged information across the network, also, the user can monitor the privacy of his data without relying on CSP information, in which case the agent moves to the source information and performs local negotiations to get the information. Work is currently going on the frame work implantation where it will be applied to a specific case study. Further research could be realized to improve and to extend the present work. Roughly speaking, we propose the following: the use of multiple mobile agents for accounting and monitoring the virtual machines at CSP, in order to improve the quality of the proposed solution and to give the user more visibility in his own information and to reinforce the trust between users and Cloud provider.